\documentstyle [12pt]{article}
\input BoxedEPS.tex
\SetOzTeXEPSFSpecial
\HideDisplacementBoxes
\begin{document}
~\\
~\\
~\\
\begin{center} 
\large {{\bf Is Room Temperature Superconductivity in Carbon Nanotubes Too 
Wonderful to Believe?}}
\end{center}
 ~\\
~\\
\mbox{\hspace{1in}Guo-meng Zhao$^{*}$}\\
~\\
\mbox{\hspace{1in}Department 
of Physics and Astronomy,}\\
\mbox{\hspace{1in}California State University at Los Angeles, }\\
\mbox{\hspace{1in}Los Angeles, CA 90032, USA}\\
~\\
~\\
{\bf ABSTRACT}: It is well known that copper-based perovskite oxides 
rightly enjoy consensus as high-temperature superconductors on the 
basis of two signatures: Meissner effect and zero resistance.  In 
contrast, I provide over twenty signatures for room temperature 
superconductivity in carbon nanotubes.  The one-dimensionality of 
the nanotubes complicates the right-of-passage for prospective 
quasi-one-dimensional superconductors.  The Meissner effect is less 
visible because the diameters of nanotubes are much smaller than the 
penetration depth.  Zero resistance is less obvious because of the quantum 
contact resistance and significant quantum phase slip, both of which 
are associated with a finite number of transverse conduction channels.  
Nonetheless, on-tube resistance at room temperature has been found to 
be indistinguishable from zero for many individual multi-walled 
nanotubes.  On the basis of more than twenty arguments, I suggest that 
carbon nanotubes deserve to be classified as room temperature 
superconductors.  The mechanism for room-temperature superconductivity 
may arise from strong electron-phonon and electron-plasmon coupling. 
~
~\\
{\bf 1~~Introduction} \vspace{0.5cm}

Finding room temperature (RT) superconductors is one of the 
most challenging problems in science.  It was theoretically shown that 
RT superconductivity cannot be realized within the conventional 
phonon-mediated pairing mechanism \cite{Mc}.  It has also been 
demonstrated that a polaronic effect can enhance superconductivity 
substantially \cite{Alex}.  But it is unclear whether RT 
superconductivity can be achieved within this mechanism.  On the other 
hand, a theoretical calculation showed that superconductivity as high 
as 500 K can be reached through a pairing interaction mediated by 
undamped acoustic plasmon modes in a quasi-one-dimensional electronic 
system \cite{Lee}.  Moreover, high-temperature superconductivity can 
occur in a multi-layer electronic system due to an attraction of 
charge carriers in the same conducting layer via exchange of virtual 
plasmons in neighboring layers \cite{Cui}.  If these theoretical 
studies are relevant, one should be able to find high-temperature 
superconductivity in quasi-one-dimensional (1D) and/or multi-layer 
systems.

Carbon nanotubes constitute a novel class of quasi-one-dimensional  materials which would offer the potential for high-temperature 
superconductivity.  The simplest single-walled nanotube (SWNT) consists of 
a single graphite sheet which is curved into a long 
cylinder, with a diameter which can be smaller than 1 nm.  
Band-structure calculations predict that carbon nanotubes have two 
types of electronic structures depending on the chirality 
\cite{Saito1,Ajiki}, which is indexed by a chiral vector $(n,m)$: $n 
-m = 3N +\nu$, where $N, n, m$ are the integers, and $\nu = 0, \pm 1$.  
The tubes with $\nu$ = 0 are metallic while the tubes with $\nu$ = 
$\pm$1 are semiconductive.  Multiwalled nanotubes (MWNTs) consist of at 
least two concentric shells which could have different chiralities.  
The MWNTs possess both quasi-one-dimensional and multi-layer electronic 
structures.  This unique quasi-one-dimensional  electronic structure in both SWNTs and MWNTs 
makes them ideal for plasmon-mediated high-temperature superconductivity.

In a series of five papers 
\cite{Zhao,Zhao2,Zhao3,Zhao4,Zhao5} resident on the cond-mat e-print archive 
since November 2001,  I provide over 
twenty arguments for RT superconductivity in carbon 
nanotubes.  In this article, I will review some of these arguments and 
provide more evidence for room-temperature superconductivity in carbon 
nanotubes.  
~\\
~\\
{\bf 2~~Electrical transport and tunneling spectra in SWNTs} 
\vspace{0.5cm}

It is well known that, for quasi-1D systems 
disorder has extremely strong effects on electrical transport 
\cite{Gia,Orignac}.  For a noninteracting 1D system, all states get 
localized in the presence of an infinitesimal random potential 
\cite{Abri}.  Interactions can modify this picture, as shown clearly 
from theoretical studies of one-chain and two-chain systems using bosonization and renormalization-group techniques 
\cite{Gia,Orignac}.  It is shown that metal-like conductivity occurs 
only for strongly attractive interactions that lead to quasi-1D s-wave 
high-temperature superconductivity \cite{Gia,Orignac}.  In contrast, for repulsive 
interactions d-wave superconductivity would occur for the pure system 
and, in the presence of disorder and/or impurities, the resistivity is 
found to decrease monotonically with increasing temperature even at 
high temperatures (semiconductor-like behavior) \cite{Orignac}.  Thus, d-wave 
superconductivity and a normal-state 
Luttinger liquid are not compatible with a metal-like resistivity in 
quasi-1D systems \cite{Orignac}.

Alternatively, the metal-like conductivity below a mean-field 
superconducting transition temperature $T_{c0}$ in quasi-1D s-wave 
superconductors can be understood in terms of quantum phase slips 
(QPS).  Indeed, QPS theories \cite{Giordano90,Zaikin} can well explain a number of 
experiments that show a large resistance and its metal-like temperature 
dependence well below $T_{c0}$ in thin 
superconducting wires \cite{Giordano90,Giordano89,Giordano91,Tinkham}. 
 
 Essentially, the phase slips at low temperatures are related to the 
macroscopic quantum tunneling (MQT), which allows the phase of the 
superconducting order parameter to fluctuate between zero and $2\pi$ 
at some points along the wire, resulting in voltage pulses.  The QPS 
tunneling rate is proportional to $\exp (-S_{QPS})$, where $S_{QPS}$ 
in clean superconductors is very close to the number of transverse 
channels $N_{ch}$ in the limit of weak damping (see below).  If the 
number of transverse channels $N_{ch}$ is small, the QPS tunneling 
rate is not negligible, leading to a non-zero resistance at low 
temperatures.  For a SWNT, $N_{ch}$ = 2, implying a large QPS tunneling 
rate and thus a finite resistance even if it is a superconductor.  For 
MWNTs with several superconducting layers adjacent to each other, $N_{ch}$  will increase substantially, 
resulting in the large suppression of the QPS if the Josephson 
coupling among the layers is strong.  If two superconducting tubes are 
closely packed together to effectively increase $N_{ch}$, one would 
find a substantially reduced  on-tube resistance at RT if the constituent 
tubes have a mean-field $T_{c0}$ well above RT.

There are thermally activated 
phase slips (TAPS) and QPS in thin superconducting wires.  TAPS occur 
via thermal activation and are dominant only at temperatures slightly 
below and close to $T_{c0}$.  At low temperatures, QPS dominate.  The 
resistance at low temperature for carbon nanotubes due to MQT is shown 
to be \cite{Zhao3}
\begin{eqnarray}\label{MQT2}
R_{MQT} =\beta_{1}\frac{h}{4e^{2}}\frac{L}{\xi}\sqrt{0.26\beta_{2}N_{m} }~\nonumber \\
\exp (-0.26\beta_{2}N_{m}), 
\end{eqnarray}
where $L$ is the length of the wire, $\xi$ is the coherence 
length, $N_{m}$ is the number of metallic chirality shells, $\beta_{1}$ 
and $\beta_{2}$ are constants, depending on the damping strength.  
When the damping increases, $\beta_{2}$ decreases.

In the limit of weak damping, $\beta_{2}$ = 7.2 (Ref.~\cite{Giordano90}). 
Therefore, from Eq.~\ref{MQT2} one can see that $S_{QPS} \simeq 2N_{m}$. For  stronger damping, $\beta_{2}$ is reduced so that $S_{QPS} < 
2N_{m}$.  Moreover, in the dirty limit, $S_{QPS}$ will be further 
reduced \cite{Zaikin} such that $S_{QPS} << 2N_{m}$.  For a SWNT, 
$N_{m}$ = 1 so that a large QPS and a nonzero resistance is expected 
below $T_{c0}$.  If several superconducting SWNTs are closely packed 
to ensure an increase in $N_{ch}$, the QPS would be substantially 
reduced.  This can explain why for a single SWNT  the on-tube resistance is appreciable 
at RT \cite{Soh,Yao} while for a bundle consisting of two strongly 
coupled SWNTs the RT resistance is very small \cite{Bachtold2000}.  For 
a MWNT with $d$ = 40 nm, a total of 27 transverse channels has been seen \cite{Pablo}.  This implies that the QPS in this single MWNT should be 
strongly suppressed.  Indeed, this MWNT has nearly zero on-tube 
resistance at RT over a length of 4 $\mu$m (Ref.~\cite{Pablo}).

A more rigorous approach quantifying the QPS in quasi-1D 
superconductors \cite{Zaikin} suggests that $S_{QPS}$ depends not only 
on $N_{ch}$ but also on the normal-state conductivity $\sigma$ ($S_{QPS} 
\propto \sigma^{2/3}$).  Therefore, one can effectively suppress the 
QPS and the resistance below $T_{c0}$ by reducing the normal-state 
resistivity.

Within the QPS theory \cite{Zaikin},  the on-tube resistance $R_{i} \propto T^{2\mu-3}$ for $k_{B}T >> 
\Phi_{\circ}I/c$, and  $R_{i} $ becomes independent of temperature and is 
proportional to $I^{2\mu-3}$ for $k_{B}T << 
\Phi_{\circ}I/c$. Here $\Phi_{\circ}$ is the quantum flux, $c$ is 
the speed of light, $I$ is the current, and $\mu$ is a quantity that characterizes 
the ground state. The zero-temperature resistance can approach zero when 
$\mu > 2$, but is finite when $\mu \leq 2$.  When $\mu < 1.5$, $R_{i}$ 
increases with decreasing temperature (semiconducting behavior), while 
for $\mu > 1.5$, $R_{i} $ decreases with decreasing temperature 
(metallic behavior).  Only if the QPS are strongly suppressed, can 
zero on-tube resistance be approached below $T_{c0}$.

For $\Phi_{\circ}I/ck_{B} << T << T_{c0}$, the four-probe or two-probe 
resistance of a superconducting wire is given by
\begin{equation}\label{R}
R (T)  = R_{0} +aT^{p}.  
\end{equation}
Here $p = 2\mu-3$, $R_{0} = R_{Q}/tN_{ch}$ for four-probe measurements, 
$R_{0} = R_{Q}/tN_{ch} +2R_{c}$ for two-probe measurements, $t$ is the 
transmission coefficient ($t \leq$ 1), $R_{Q}$ = $h/2e^{2}$ =12.9 
k$\Omega$ is the resistance quantum, and $R_{c}$ is the contact 
resistance.

\begin{figure}[htb]
\ForceWidth{7cm}
\leftline{~~~\BoxedEPSF{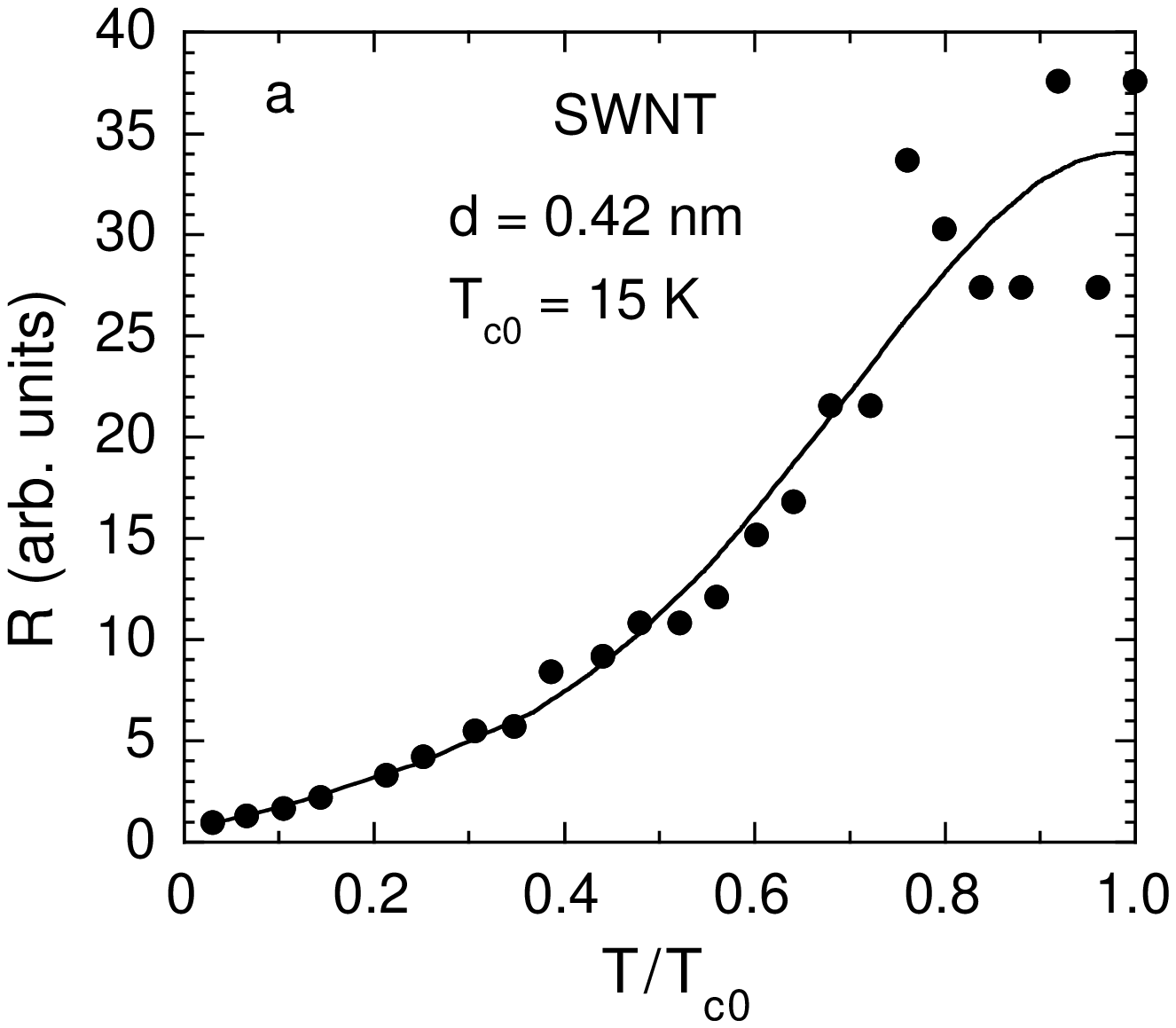}} 
\ForceWidth{7cm} 
\vspace{-6.1cm}
\rightline{\BoxedEPSF{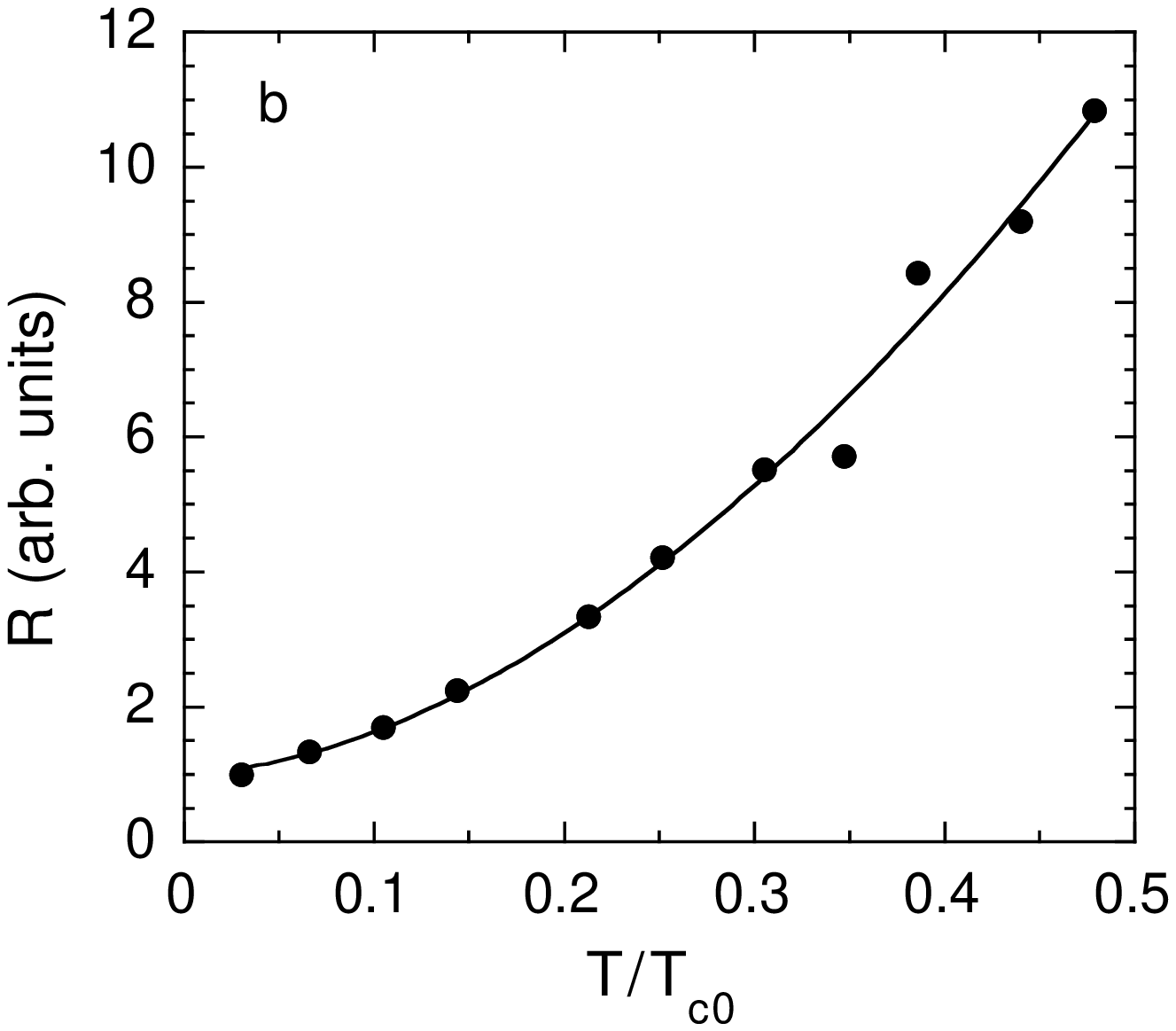}~~~} 
\caption [~]{a) The resistance as a 
function of $T/T_{c0}$ for the smallest diameter SWNTs with $d$ = 0.42 
nm.  The data are extracted from Ref.~\cite{Tang}.  b) The temperature 
dependence of the resistance below 0.5$T_{c0}$.  }
\label{1Q}
\end{figure}

Now we compare the QPS theory with the resistance data of SWNTs that are 
known to exhibit quasi-1D superconductivity.  For the smallest 
diameter SWNTs with $d$ = 0.42~nm,  $T_{c0}$ was found to be about 15 K 
\cite{Tang}.  In Fig.~\ref{1Q}a, we plot the two-probe resistance as a 
function of $T/T_{c0}$ for this SWNT.  It is apparent that the resistance increases more rapidly above 
0.5$T_{c0}$ and flattens out towards $T_{c0}$.  
From Fig.~1a, we also see that the normal-state 
resistance $R_{N}$ at $T_{c0}$ is larger than the zero-temperature 
resistance $R_{0}$ by a factor of 34.  Since $R_{0} \geq 
R_{Q}/N_{ch}$, then $R_{N} \geq 34R_{Q}/N_{ch}$.  With $N_{ch} $= 8 
per tube (Ref.~\cite{Zhao4}), we obtain $R_{N} \geq 34R_{Q}/8$ = 55 
k$\Omega$.  Because the contact distance is about 50~nm \cite{Tang}, the 
normal-state resistance per tube per unit length is larger than 1100 
k$\Omega$/$\mu$m .  This suggests a very short mean free path 
($<$ 15 \AA) that makes ballistic transport impossible.

As demonstrated in Fig.~\ref{1Q}b, below 0.5$T_{c0}$ the temperature 
dependence of the resistance can be well fitted by Eq.~\ref{R}. From 
the fit, we find that $p$ = 1.77$\pm$0.18.  This suggests that the QPS 
theory can indeed explain the quasi-1D superconductivity in SWNTs.  
Using $p$ = 1.77 and $p = 2\mu -3$, we obtain $\mu$ = 2.4 $>$ 2, 
implying zero on-tube resistance at zero temperature.

\begin{figure}[htb]
\ForceWidth{7.0cm}
\leftline{~~~~~~~~~\BoxedEPSF{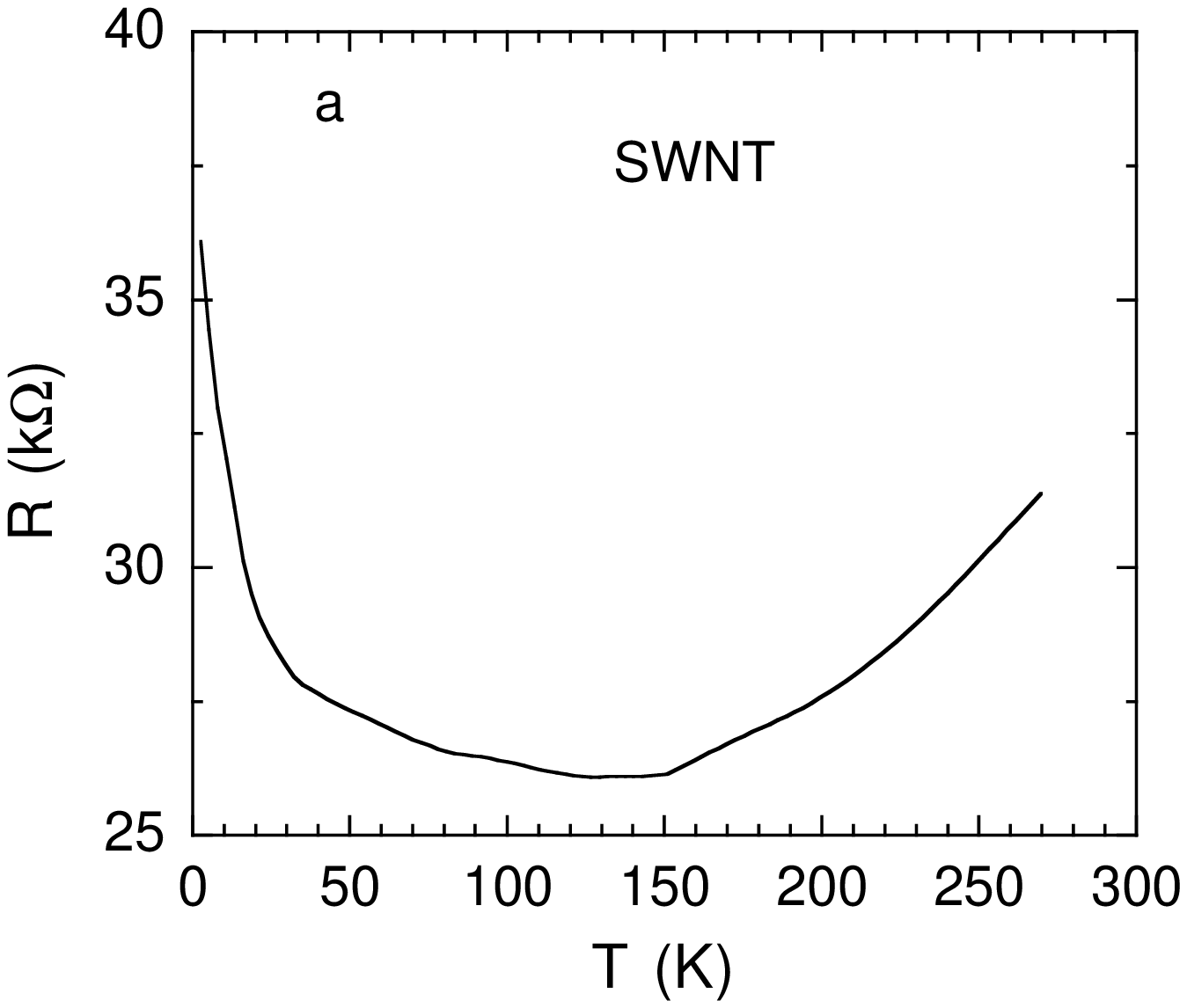}} 
\ForceWidth{5.0cm} 
\vspace{-6.0cm}
\rightline{\BoxedEPSF{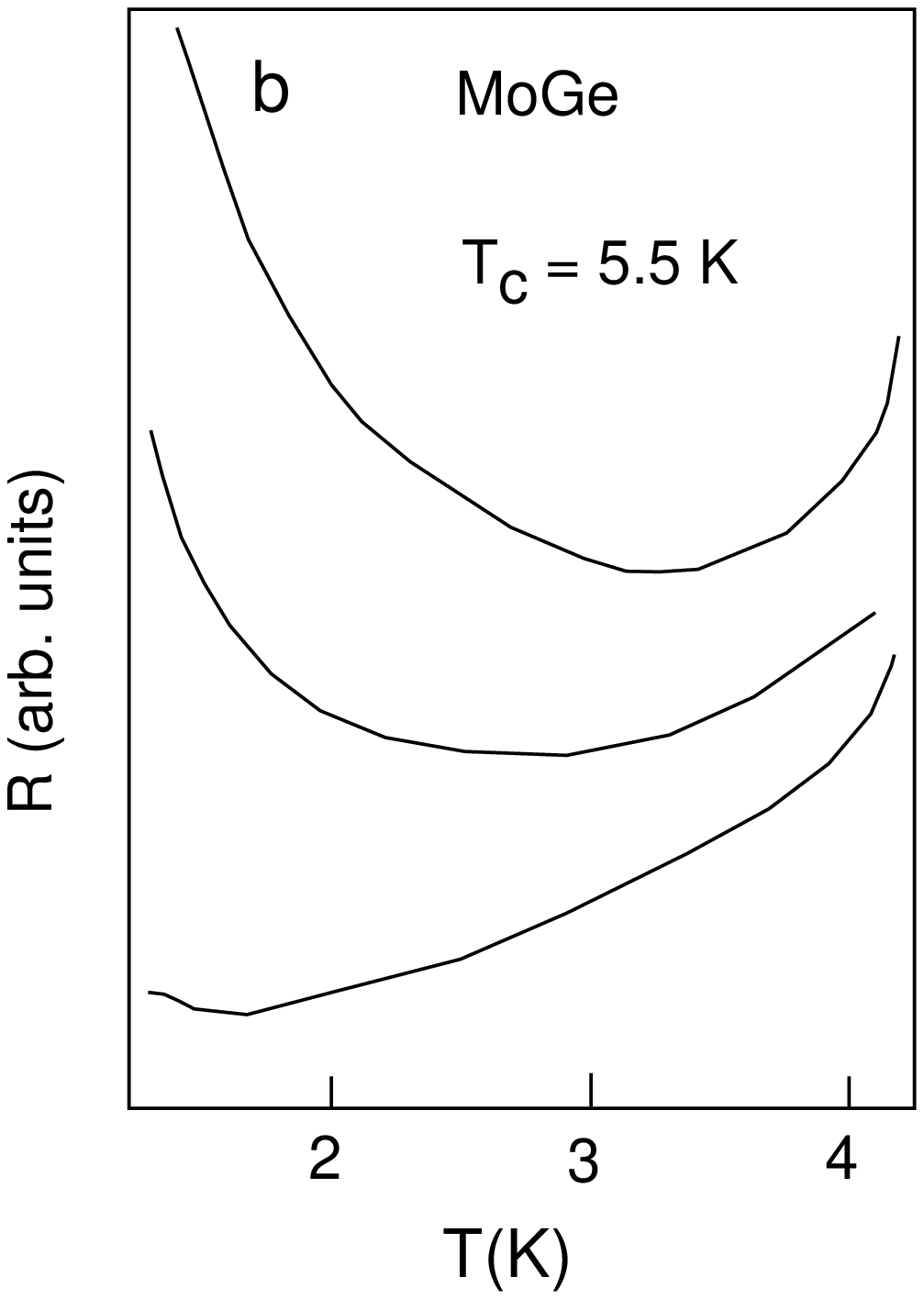}~~~~~~~~~~~} 
\caption [~]{a) Temperature dependence 
of the resistance for a SWNT.  The data are extracted from 
Ref.\cite{Soh}.  b) Temperature dependence of the resistance for three 
ultrathin MoGe wires.  The curves are smoothed from the original plot 
of Ref.~\cite{Tinkham}.}
\label{2Q}
\end{figure}

In Fig.~\ref{2Q}a, we plot the temperature dependence of the resistance for a 
single SWNT. It is interesting that the temperature dependence of the 
resistance in this SWNT is similar to that found for ultrathin wires of MoGe 
superconductors \cite{Tinkham}, a facsimile of which is reproduced in 
Fig.~\ref{2Q}b.  By comparing Fig.~\ref{2Q}a and Fig.~\ref{2Q}b, one 
might infer that the mean-field $T_{c0}$ of this nanotube is well 
above 270 K.

We can fit the data of Fig.~\ref{2Q}a above a  resistance-minimum temperature 
($\sim$150 K) with Eq.~\ref{R}.  This will result in $p$ $>$ 1, which 
is inconsistent with electrical transport for a Luttinger liquid (LL).  
Phonon backscattering and/or electron umklapp scattering in a LL lead to 
a semiconductor-like resistivity at low temperatures and a power-law 
resistivity with $p$ $\simeq$ 0.6 at high temperatures 
\cite{Komnik,Kane}.  Disorder gives rise to a high-temperature 
power-law resistivity with $p$ $\simeq$ $-$0.4 (Ref.~\cite{Kane}).  
Any combinations of these power laws predicted from the LL are not 
compatible with the data shown in Fig.~\ref{2Q}a.  Further, recent tunneling 
experiments have excluded LL behavior \cite{Tark}.  Moreover, the 
data in Fig.~\ref{2Q}a also rule out quasi-1D d-wave superconductivity 
because quasi-1D d-wave superconductors behave like insulators at any 
temperatures \cite{Orignac}.
\begin{figure}[htb]
\ForceWidth{7cm}
\centerline{\BoxedEPSF{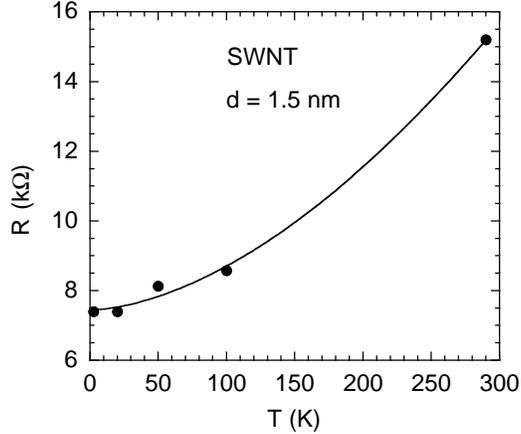}}
\caption [~]{Temperature dependence of the resistance at zero gate 
voltage for a single-walled nanotube with $d$ = 1.5 nm.  The data are 
extracted from Fig.~1a of Ref.~\cite{Kong}. }
\label{3Q}
\end{figure}

Fig.~\ref{3Q} shows the 
temperature dependence of the resistance at zero gate voltage for a 
single-walled nanotube with $d$ = 1.5 nm.  The distance between the two contacts is about 
200 nm and the contacts are nearly ideal with the transmission 
probability of about 1 \cite{Kong}.  It is remarkable that the 
temperature dependence of the resistance can be fitted by Eq.~\ref{R} 
with $p$ = 1.71$\pm$0.23.  This exponent is very close to the one 
found for the smallest SWNTs with $T_{c0}$ = 15 K (see Fig.~\ref{1Q}b).  
Further, the on-tube resistance at zero temperature is found to be nearly zero 
\cite{Kong}.  These results can be only explained by the QPS theory 
\cite{Zaikin}.  Using $p$ = 1.71 and $p = 2\mu -3$, we get $\mu$ = 
2.35 $>$ 2.  This implies a negligible on-tube resistance at zero 
temperature according to the QPS theory \cite{Zaikin}, in agreement 
with experiment \cite{Kong}.

 For a longer tube where the distance between the two contacts is 
 about 800~nm, the resistance at zero gate voltage is independent of 
 temperature below 270 K (Ref.~\cite{Kong}), i.e., $p$ $\simeq$ 0.  It was also shown that 
 \cite{Kong} underdoping leads to $p$ $<$ 0 while overdoping gives rise 
 to $p$ $>$ 0.  This doping dependence of $p$ is only consistent with 
 the QPS theory.  Within the QPS theory \cite{Zaikin}, $\mu$~$\propto 
 r/\lambda_{L} $ $\propto \sqrt{n}$ (where $\lambda_{L}$ is the London 
 penetration depth and $n$ is the carrier density).  It is clear that 
 $\mu$ increases with $n$ such that $p$ can cross over from a negative 
 value at low doping to a positive one at high doping, in agreement 
 with experiment \cite{Kong}.

Now we discuss tunneling spectra. From a  single-particle tunneling spectrum 
obtained through two high-resistance contacts (see Fig.~1b of 
Ref.~\cite{Yao}), we can clearly see a pseudo-gap feature which 
appears at an energy of about 220 meV.  The pseudo-gap feature could be 
related to the superconducting gap.  Considering the broadening of the 
gap feature due to large QPS and the double tunneling junctions in 
series, we estimate the superconducting gap $\Delta (0)$ to be about 100 meV.  The scanning tunneling microscopy and spectroscopy 
\cite{Wildoer} on individual single-walled nanotubes also show  pseudo-gap features with $\Delta (0) \simeq$ 100 meV in doped metallic 
SWNTs.  Using $k_{B}T_{c0} = \Delta (0)/1.76$, we find $T_{c0} \simeq$ 
660 K.
\begin{figure}[htb]
\ForceWidth{7cm}
\centerline{\BoxedEPSF{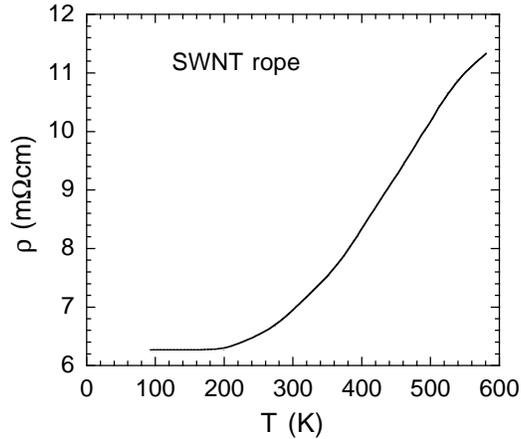}}
	\caption [~]{Temperature dependence of the resistivity for a 
SWNT rope with $T_{c0}$ $\simeq$ 640 K. The data are extracted from Ref.~\cite{Nature}.}
\label{4Q}
\end{figure}

 This gap value deduced from the tunneling spectrum is in good 
agreement with electrical transport data for a SWNT rope (see 
Fig.~\ref{4Q}).  Below 200 K the resistivity is nearly temperature 
independent while above 200 K the resistivity increases suddenly and 
starts to flatten out above 550 K.  By comparing with Fig.~\ref{1Q}a, 
we find that this temperature dependence of the resistance agrees with 
quasi-1D superconductivity with a $T_{c0}$ $\simeq$ 640 K.  If we 
assume that one-third of the tubes have metallic chiralities (MC) and 
exhibit superconductivity and that the on-tube resistance is 
negligible below 200 K, we find that the normal-state on-tube 
resistivity for the MC tubes is about 1.6 m$\Omega$cm.

\begin{figure}[ht]
\ForceWidth{6.2cm}
\centerline{\BoxedEPSF{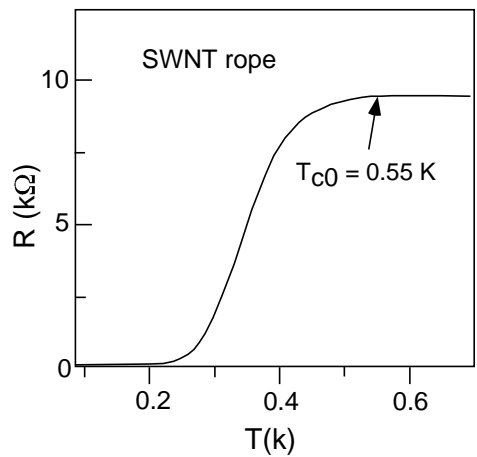}}
\caption [~]{The 
temperature dependence of the resistance for a SWNT rope with 
$T_{c0}$ = 0.55 K. The figure 
is reproduced from Ref.~\cite{Kociak}}
\label{5Q}
\end{figure}

 The electrical transport and single-particle tunneling spectra 
thus consistently suggest that $T_{c0}$ in SWNTs can be higher than 500 K.  
Nevertheless, a much lower $T_{c0}$ 
was observed in a SWNT rope \cite{Kociak}.  Fig.~\ref{5Q}  shows temperature 
dependence of the resistance for the SWNT rope.  One can see that the resistance 
starts to drop below about 0.55 K.  The data agree with quasi-1D 
superconductivity with $T_{c0} \simeq$ 0.55 K.  It is remarkable that 
the relative transition width $\Delta T/T_{c0}$ for this SWNT rope is 
similar to that for another SWNT rope with $T_{c0}$ $\simeq$ 640 K 
(see Fig.~\ref{4Q} ).  With $R_{0}$ = 74 $\Omega$ and assuming perfect 
contacts, one finds \cite{Kociak} that the normal-state resistance per 
tube per unit length is 830 k$\Omega$/$\mu$m which corresponds to the 
normal-state resistivity of 128 $\mu\Omega$cm and a mean free path of 
78~\AA.  We should mention that very low superconductivity in this 
SWNT rope may be due to the fact that the tubes are very lightly 
doped.  This is in agreement with the tunneling spectrum which shows 
no pseudo-gap feature at 5 K in an undoped or very lightly doped 
armchair tube \cite{Lieber}.  The strong doping dependence of $T_{c0}$ 
is consistent with the superconducting mechanism based on 
electron-plasmon coupling \cite{Lee}.  ~\\
~\\
{\bf 3~~Raman scattering in a SWNT rope} 
\vspace{0.5cm}

It is known that Raman scattering has provided essential information 
about the electron-phonon coupling and the electronic pair excitation 
energy in the high-$T_{c}$ cuprate superconductors 
\cite{Krantz,Cardona,Ham}.  The anomalous temperature-dependent 
broadening of the Raman active $B_{1g}$-like mode of 90 K 
superconductors RBa$_{2}$Cu$_{3}$O$_{7-y}$ (R is a rare-earth element) 
allows one to precisely determine the superconducting gap \cite{Cardona}.  
The pronounced 
softening observed only for the $B_{1g}$ mode is due to the fact that  
the phonon energy of the $B_{1g}$ mode is very close to 2$\Delta (0)$ and 
the mode is strongly coupled to electrons \cite{Cardona,ZZ}. 

\begin{figure}[htb]
\ForceWidth{7cm}
\leftline{~~~~~~\BoxedEPSF{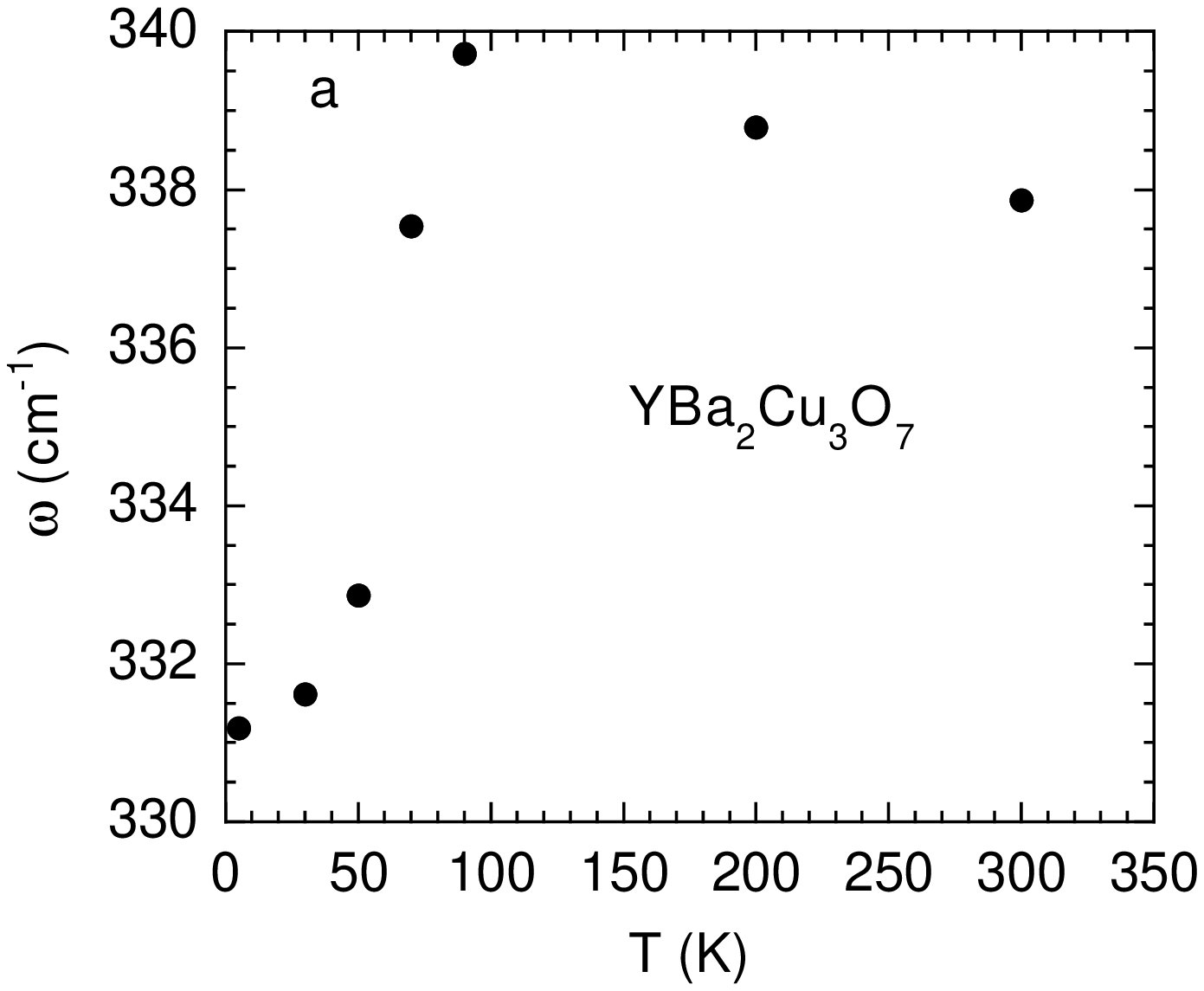}} 
\ForceWidth{6.8cm} 
\vspace{-5.7cm}
\rightline{\BoxedEPSF{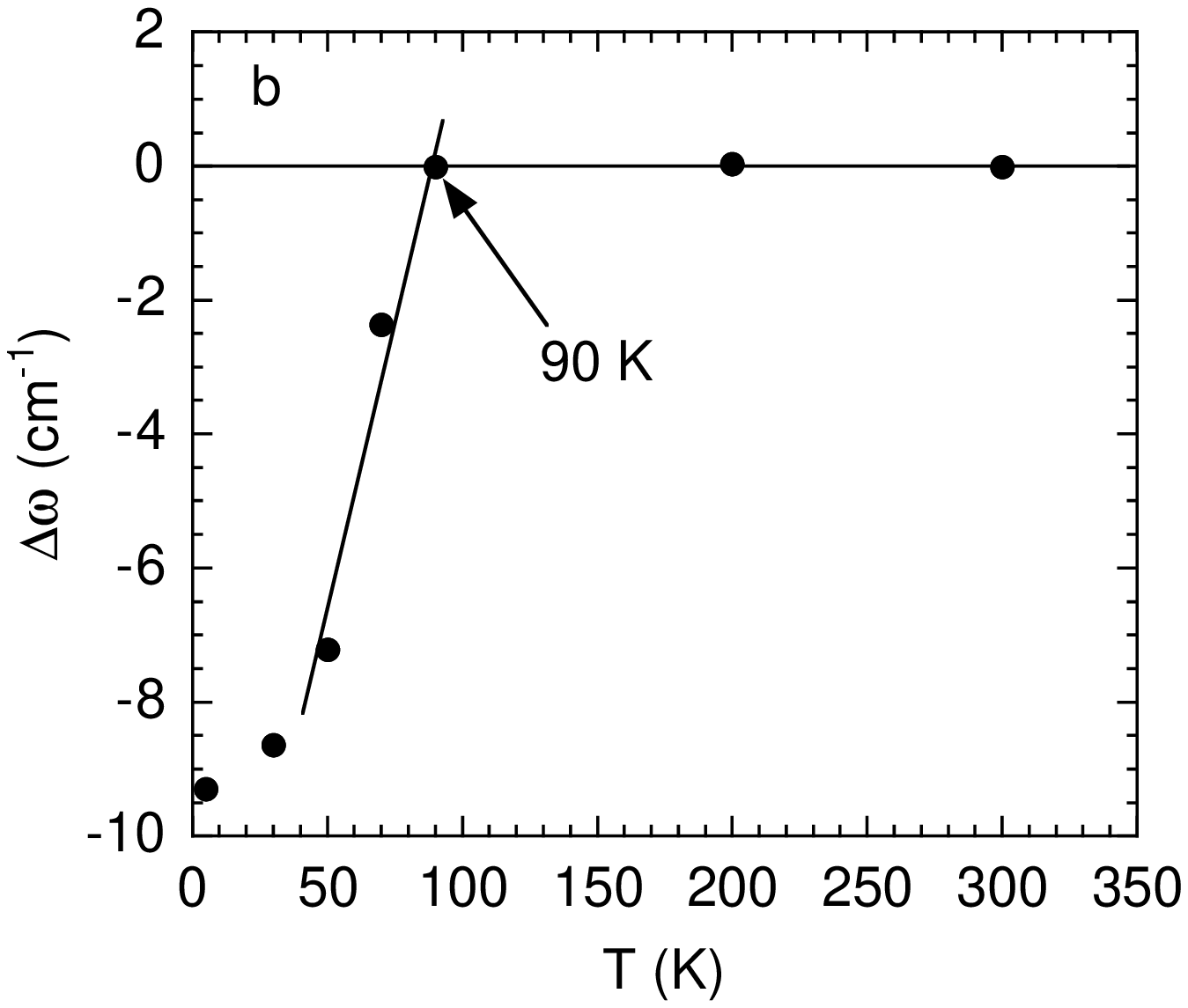}~~~~~~} 
\caption [~]{a) Temperature dependence 
of the frequency for the Raman-active $B_{1g}$ mode of a 90 K 
superconductor YBa$_{2}$Cu$_{3}$O$_{7-y}$.  The data are extracted 
from Ref.\cite{Krantz}.  b) The difference between the measured 
frequency and the linearly fitted curve above $T_{c}$.}
\label{1R}
\end{figure}

The temperature dependence of the frequency for the 
Raman-active $B_{1g}$ mode of a 90 K superconductor 
YBa$_{2}$Cu$_{3}$O$_{7-y}$ is shown in Fig.~\ref{1R}a.  It is apparent 
that the frequency decreases linearly with increasing temperature 
above $T_{c}$ and that the mode starts to soften below about 0.95$T_{c}$.  The 
temperature dependence of the frequency above $T_{c}$ is caused by 
thermal expansion.  The temperature dependence of the frequency will 
become more pronounced at higher temperatures since the magnitude of 
the slope $-d\ln\omega/dT$ is essentially proportional to the lattice 
heat capacity that increases monotonically with temperature.  The 
significant softening of the mode below $T_{c}$ occurs only if the 
energy of the Raman mode is very close to 2$\Delta (0)$ and the 
electron-phonon coupling is substantial \cite{ZZ}, as it is the case 
in the 90 K superconductor YBa$_{2}$Cu$_{3}$O$_{7-y}$ 
\cite{Krantz,Cardona,Ham}.  In order to see more clearly the softening 
of the mode, we show in Fig.~1b the difference of the measured 
frequency and the linearly fitted curve above $T_{c}$.  It is clear 
that the softening starts at about 89 K $\simeq$ 0.95$T_{c}$.

\begin{figure}[htb]
\ForceWidth{7cm}
\leftline{~~~~~\BoxedEPSF{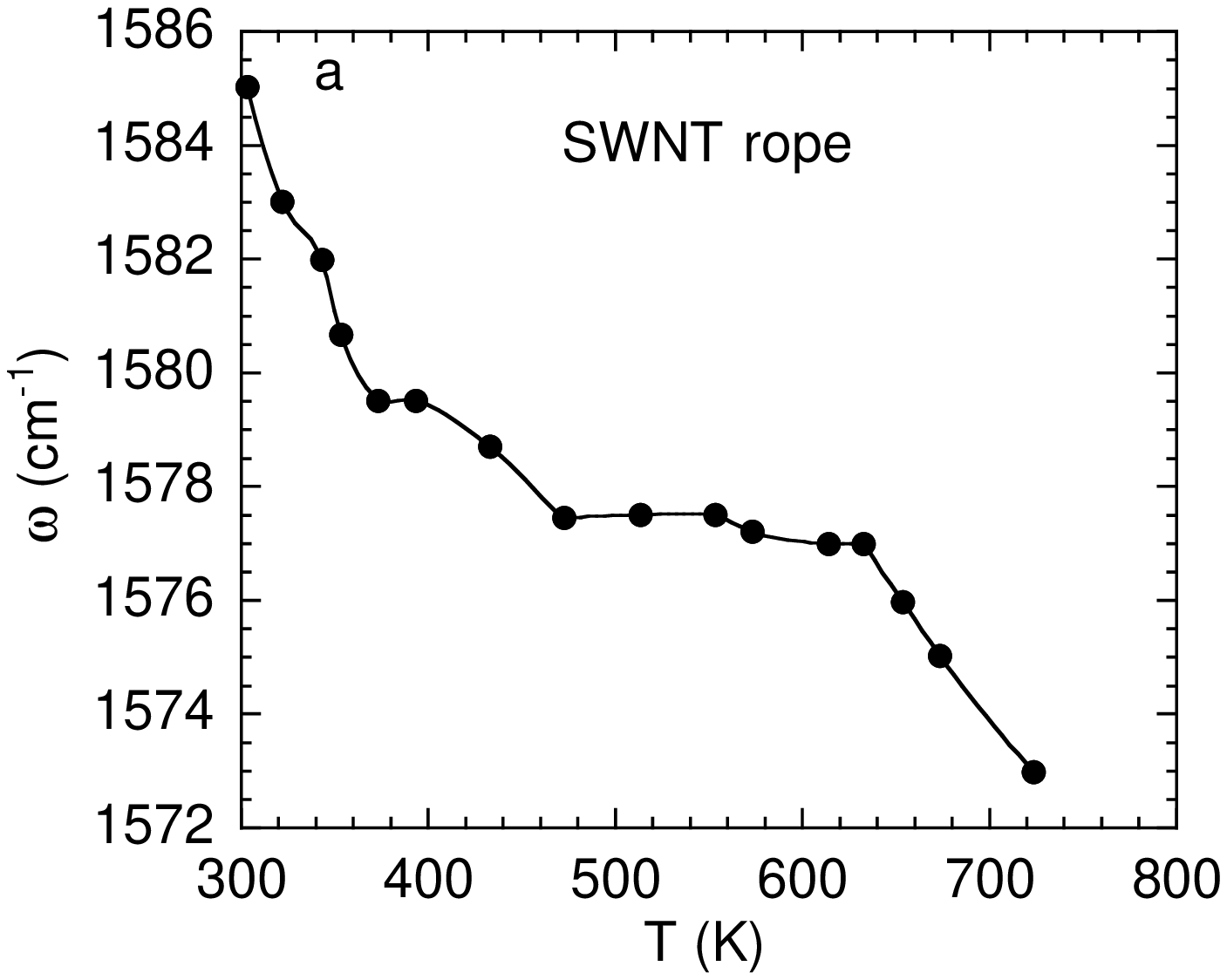}} 
\ForceWidth{6.8cm}
\vspace{-5.5cm}
 \rightline{\BoxedEPSF{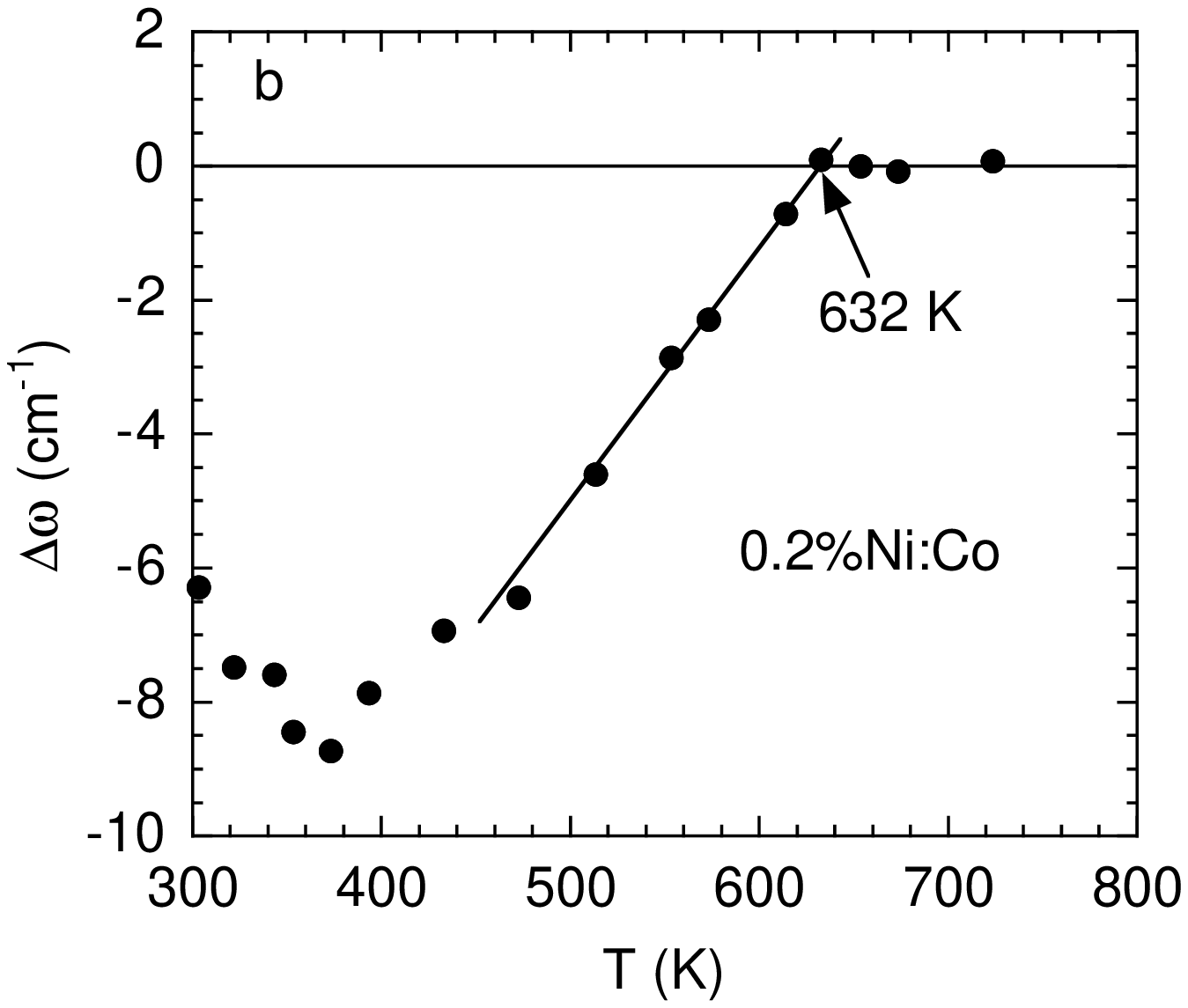}~~~~~} 
 \caption [~]{a) Temperature 
dependence of the frequency for the Raman active $G$-band of a SWNT 
rope.  The data are from Ref.~\cite{Walter}.  b) The difference of the 
measured frequency and the linearly fitted curve above the kink 
temperatures (see text).}
\label{2R}
\end{figure}

Fig.~\ref{2R}a shows the temperature dependence of the frequency for the 
Raman active $G$-band of a SWNT rope.   It is 
remarkable that the frequency data show a clear tendency of softening 
below about 630 K. Above 630 K, the frequency decreases linearly with 
increasing temperature, which is similar to the behavior in 
YBa$_{2}$Cu$_{3}$O$_{7-y}$ (Fig.~\ref{1R}a).

 In order to see more clearly the softening of the mode, we 
show in Fig.~\ref{2R}b the difference between the measured frequency 
and the linearly fitted curve above the kink temperature ($\sim$630 K).  It is 
striking that the result shown in Fig.~\ref{2R}b is 
similar to that shown in Fig.~\ref{1R}b.  This suggests that the 
softening of the Raman active $G$-band in the SWNTs may have the same 
microscopic origin as the softening of the Raman active $B_{1g}$ mode 
in YBa$_{2}$Cu$_{3}$O$_{7-y}$.  This explanation is plausible only if 
the phonon energy of the $G$-band is very close to 2$\Delta (0)$.  We are 
fortunate that  the phonon energy of the $G$-band is 197 meV, very close to 
2$\Delta (0)$ = 200 meV deduced from the tunneling spectrum and the 
electrical transport experiment \cite{Zhao3}.  Therefore, it is very 
likely that the softening of the Raman mode in the SWNTs is related to 
a superconducting phase transition.

From Fig.~\ref{2R}, we can clearly see that the softening starts at 
about 632 K.  Using the fact that the softening starts at 
0.95$T_{c0}$ (Ref.~\cite{ZZ}), we can assign the mean-field transition 
temperature $T_{c0}$ = 665 K.

It is interesting to note that there is a clear  minimum at $T^{*}$ = 
370 K = 0.57$T_{c0}$ in Fig.~\ref{2R}b.  It is remarkable that such a 
minimum is also seen at about 0.6$T_{c0}$ in a calculated curve for a 
superconductor  (see Fig.~8 of Ref.~\cite{ZZ} ).  This shallow minimum 
in the frequency shift is related to a sharp minimum in the real part 
of the polarization $\Pi(\omega,T)$, which occurs at 
$\hbar\omega$/$2\Delta (T^{*})$ = 1 for weak coupling \cite{ZZ,ZZ2}.  
This quantitative consistency provides strong evidence that the 
softening of the Raman active mode in the SWNTs is indeed associated 
with a superconducting phase transition.

From the temperature dependence of the BCS gap \cite{book}, we find 
that $\Delta (T^{*})$ = $\Delta (0.57T_{c0})$ = 0.93$\Delta (0)$.  
Using $\hbar\omega$/$2\Delta (T^{*})$ = 1 and $\hbar\omega$ = 197 
meV, we finally obtain $\Delta (0)$ = 105 meV, which is in excellent 
agreement with the value deduced from the tunneling spectrum 
\cite{Zhao3}.  Then we calculate 2$\Delta (0)/k_{B}T_{c0}$ = 3.66, 
which is very close to that expected from the weak-coupling BCS theory.  This 
value is also in excellent agreement with the prediction based on the 
plasmon-mediated pairing mechanism \cite{Lee}.  ~\\
~\\
{\bf 4~~Electrical transport and tunneling spectrum in MWNTs} 
\vspace{0.5cm}

Subtle evidence for superconductivity above RT can be found in  MWNT ropes 
\cite{Zhao}.  Further, a negligible on-tube resistance ($\sim$86 
$\Omega$/$\mu$m) has been observed at RT in the majority of individual 
MWNTs \cite{Frank,Heer}.  If we use the average diameter of 15 nm for 
the MWNTs, we find that the average room-temperature resistivity is 
1.5 $\mu$$\Omega$cm, which is about three orders of magnitude smaller 
than the normal state resistivity for SWNTs.  In order to explain such 
a small on-tube resistivity at RT, one must assume that these MWNTs 
have $T_{c0}$'s much higher than 300 K and QPS are significantly 
suppressed.
\begin{figure}[htb]
\ForceWidth{7cm}
\centerline{\BoxedEPSF{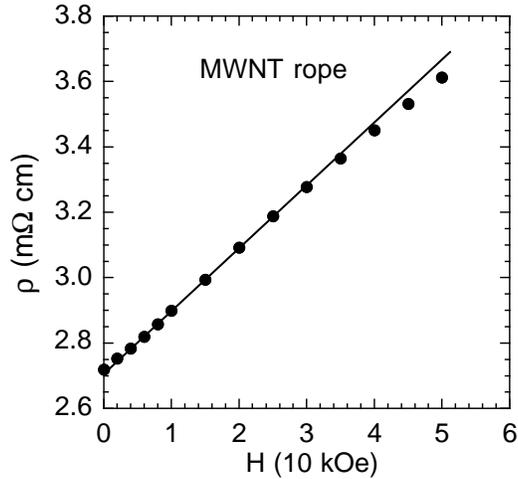}}
\caption [~]{The magnetic field dependence of the resistivity at 298 K 
for a MWNT rope.}
\label{MWR}
\end{figure}

In Fig.~\ref{MWR}, we show the magnetic field dependence of the 
resistivity at 298 K for a MWNT rope.  For our MWNT ropes, we find that the 
average weight density is about 40$\%$ of the
microscopic weight density (2.17 g/cm$^{3}$) \cite{Qian}.  The resistivity is 
calculated with this porosity correction.  From Fig.~\ref{MWR}, we can 
clearly see that the zero-field resistivity for the rope is about 
three orders of magnitude larger than the on-tube resistivity.  This 
suggests that the room-temperature resistivity for this rope is 
dominated by the contact resistivity between tubes.  Since there 
should be no magnetoresistance (MR) effect for the contact resistance, 
the result shown in Fig.~\ref{MWR} implies a very large on-tube MR 
effect, i.e., $\Delta \rho_{i}/\rho_{i}$ $\simeq$ 60000$\%$
at $H$ = 50 kOe (using $\rho_{i}$ =1.5 $\mu$$\Omega$cm).  Such a huge 
on-tube MR effect is hard to explain without invoking superconductivity.

Since there is a negligible positive MR effect for physically 
separated  MWNTs up to 180 K (Ref.~\cite{Baum}), the observed large 
positive MR effect in physically coupled  MWNT rope may be associated 
with the coupling of the tubes.  However, no positive MR is observed 
up to 225 K for physically coupled SWNT ropes \cite{Kim}.  This 
suggests that the tube coupling alone is not sufficient to produce a 
huge positive MR effect.  It is clear that the combined results cannot 
be explained if MWNTs were not RT superconductors.   It is well known that the 
motion of vortices in a vortex-liquid state can cause resistivity that is proportional to the magnetic field \cite{book}.  
When the sizes of all individual MWNTs are not large enough to trap 
any vortices \cite{Millis}, there will be no positive MR effect.  This 
can account for the negligible positive MR effect for physically 
separated MWNTs.  When individual MWNTs are closely packed into 
bundles, the sizes of Josephson coupled bundles are large enough to 
trap vortices so that a large positive MR can be observed in the 
vortex-liquid state, in agreement with the result shown in 
Fig.~\ref{MWR}.  No positive MR effect in physically coupled SWNT 
ropes suggests that the SWNTs are either nonsuperconducting or the sizes of 
Josephson coupled superconducting bundles are not large enough to trap 
any vortices.
\begin{figure}[htb]
\ForceWidth{6cm}
\centerline{\BoxedEPSF{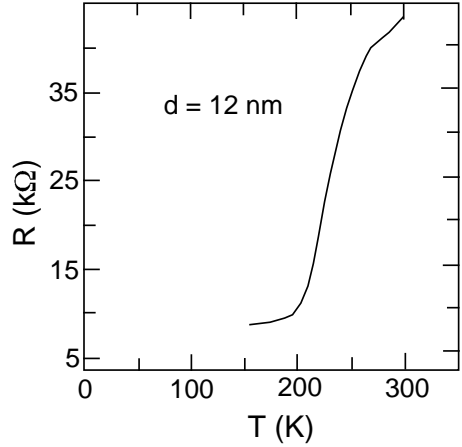}}
\caption [~]{The 
temperature dependence of the resistance for a single MWNT 
with $d$ $\simeq$ 12 nm, which is reproduced from Ref.~\cite{Eb}. The 
data are consistent with quasi-1D superconductivity with $T_{c0}$ $\simeq$ 
400 K.}
\label{MWR1}
\end{figure}

If the positive MR effect is indeed caused by 
RT superconductivity, the normal-state resistivity at 
room-temperature should be larger than $\rho (50 kOe) -\rho (0) =$ 0.9~m$\Omega$cm (see Fig.~\ref{MWR}).  This is because the resistivity is not 
completely saturated at $H$ = 50 kOe.  This normal-state resistivity 
appears to be similar to the one for a SWNT rope ($\sim$1.6~m$\Omega$cm).  From the field dependence of the resistivity (see 
Fig.~\ref{MWR}), one can also see that the melting field at 
room-temperature is very small.

 In Fig.~\ref{MWR1} we plot the temperature dependence of the resistance for a 
 single MWNT with $d$ $\simeq$ 12 nm. The contact distance is 0.5 $\mu$m.  It is striking 
 that the temperature dependence of the resistance for this MWNT is 
 similar to that (see Fig.~\ref{5Q}) for a SWNT rope consisting of about 90 
 superconducting SWNTs.  Below 200 K, the resistance is nearly 
 temperature independent, which suggests that the on-tube resistance 
 may be negligible below 200 K.  By comparing Fig.~\ref{MWR1} with 
 Fig.~\ref{5Q}, we may infer that $T_{c0}$ $\simeq$ 400 K.  Assuming zero 
 on-tube resistance below 200 K, we find that the normal-state 
 resistance per unit length at 400 K is about 66 k$\Omega$/$\mu$m.  
 This corresponds to the normal-state resistivity of 0.75 m$\Omega$cm. 
 This normal-state resistivity is very close to that deduced from 
 Fig.~\ref{MWR} ($\geq$0.9 m$\Omega$cm), and also agrees with the value 
 (1.6 m$\Omega$cm)  for a SWNT rope (see Fig.~\ref{4Q}).

From Fig.~\ref{MWR1}, we also see negligible on-tube resistance below 200 K $\simeq$ 0.5$T_{c0}$ for this individual MWNT.  By analogy, the 
negligible on-tube resistance at RT observed in the 
majority of MWNTs \cite{Heer} indicates that most MWNTs have $T_{c0}$'s higher 
than 600 K in agreement with the high-temperature resistance data  \cite{Zhao}.
\begin{figure}[htb]
\ForceWidth{6.5cm}
\centerline{\BoxedEPSF{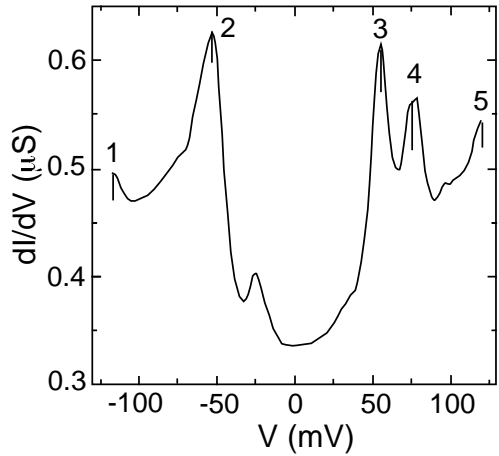}}
\caption [~]{Single-particle tunneling spectrum  for a single MWNT with $d$ 
$\simeq$ 17 nm, which is reproduced from Ref.~\cite{Sheo}.  }
\label{MWT}
\end{figure}

Fig.~\ref{MWT} shows single-particle tunneling spectrum for a single MWNT 
with $d$ =17~nm.    It 
is apparent that both the intensity and the peak position in this 
spectrum are inconsistent with the normal-state density of states for 
any types of metallic-chirality tubes \cite{Kwon,Sanvito}.  On the 
other hand, this spectrum is very similar to the one for a 
superconductor with a superconducting gap of about 54.4 meV.  This gap 
size corresponds to $T_{c0}$ = 360 K according to the BCS relation.

We can quantitatively explain this spectrum by invoking superconductivity.  The peak positions denoted by 1, 2, 3, 4, 5 are 
located at $<-$117 mV, $-$54.4 mV, 54.4 mV, 76.5 mV, and $\sim$120 mV, 
respectively.  Below $T_{c0}$, the Van Hove singularities are located 
at $E_{VH}$ = $\pm$$\sqrt{[\Delta 
(0)]^{2}+(\epsilon_{VH}-\epsilon_{F})^{2}} + \epsilon_{F}$.  Here the 
sign in the prefactor of the square root is the same as the sign of 
$\epsilon_{VH}-\epsilon_{F}$.  In order to get a consistent 
explanation to the spectrum, we assume that the tube is hole doped 
(i.e., $\epsilon_{F}$ $<$ 0) and that the positive voltage corresponds 
to the negative electron energy.

It is interesting that peak 2 is 
broader than peak 3.  This indicates that peak 2 corresponds to two 
overlapped peaks: one is a superconducting quasi-particle peak and 
another is a Van Hove singularity located slightly above the Fermi 
level.  Peak 3 corresponds to another superconducting quasiparticle 
peak.  From the positions of peak 2, 3, and 4, we deduce that $\Delta 
(0)$ = 54.4 meV, $\epsilon_{VH1}$ = $\pm$54.4 meV, $\epsilon_{VH2}$ = 
$\pm$108.5 meV $\simeq$ 2$\epsilon_{VH1}$, and $\epsilon_{F}$ = $-$55.0 meV.  The positions of Van 
Hove singularities agree with the theoretical prediction for an armchair 
tube \cite{Mintmire}.  With the values of $\epsilon_{VH1}$, $\epsilon_{VH2}$ and $\Delta 
(0)$, we can predict the other peak positions.  For example, we 
predict $V_{1}$=$-$121.6 mV and $V_{5}$= +121.6 mV--in excellent 
agreement with the measured results.  Using $\epsilon_{VH1}$ = 54.4 
meV, $d$ = 17 nm, and the relation $\epsilon_{VH1} = 
3a_{C-C}\gamma_{\circ}/d$ (Ref.~\cite{Mintmire}), we find that 
$\gamma_{\circ}$ = 2.2 eV, in quantitative agreement with another 
independent measurement \cite{Collins}.

Since the Fermi level is right on the first Van Hove singularity in 
this MWNT, the large enhancement in the density of states due to 
the Van Hove singularity leads to a large increase in the condensation 
energy, and thus a large suppression of QPS \cite{Zhao3}.  The 
observed sharp quasiparticle peak at the gap edge should be related to 
the substantially reduced QPS.  The subgap features in the spectrum may be 
related to QPS or gap anisotropy.  The much lower $T_{c0}$ in the 
outermost layer of this MWNT is due to the fact that electron-plasmon 
coupling is significantly reduced when the second 
subband is crossed \cite{Lee}.
~\\
~\\
{\bf 5~~Density of states in MWNTs} 
\vspace{0.5cm} 

As discussed above, there are metallic and semiconducting chirality 
shells that are nested to form MWNTs.  Statistically, there should be 
one-third metallic chirality shells.  However, electrical transport 
measurements on physically separated MWNTs indicate that about 80$\%$ 
of the outermost shells have metallic chiralities \cite{Sheo}.  This 
implies that about 80$\%$ of the shells
in MWNTs should have
metallic
chiralities.  This conclusion is supported by an independent measurement 
of electron spin susceptibility using ESR technique \cite{Chau}. 
\begin{figure}[htb]
\ForceWidth{7cm}
\centerline{\BoxedEPSF{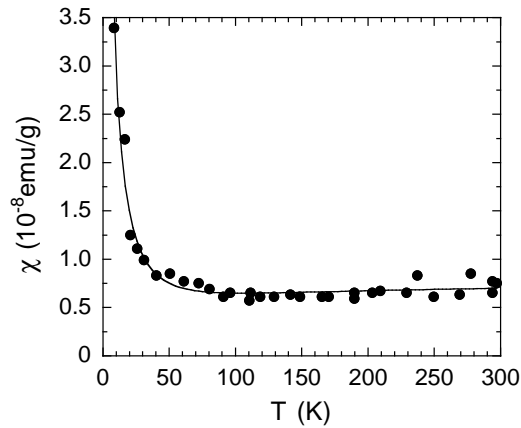}}
\caption [~]{Electron spin susceptibility for a physically separated 
MWNT film.  The data are extracted from 
Ref.~\cite{Chau}.  }
\label{ESR}
\end{figure} 

Fig.~\ref{ESR} shows the temperature dependence of the spin 
susceptibility for physically separated MWNTs.  It is interesting that 
below about 30 K the spin susceptibility is dominated by a Curie term 
($\propto$ $1/T$).  We can attribute this term to the localized 
electrons in semiconducting chirality shells because doped 
semiconducting chirality tubes are shown to be insulating \cite{Paul}.  
If the remaining metallic chirality shells are room-temperature 
superconductors with a minimum superconducting gap $\Delta_{min}$, the 
temperature dependence of the density of states should be proportional 
to $\exp(-\Delta_{min}/k_{B}T)$.  The total susceptibility is then 
given by
\begin{equation}\label{DOS}
 \chi_{s} = \frac{C}{T}+\chi_{\infty}\exp(-\Delta_{min}/k_{B}T).
\end{equation}

The solid line in Fig.~\ref{ESR} is the fitted curve of Eq.~\ref{DOS}.  
One can see that the fit is good.  The fitting parameters are  
$\chi_{\infty}$ = (7.8$\pm$0.8)$\times$10$^{-9}$ emu/g, $\Delta_{min}$ 
= 6.9$\pm$1.5 meV, and $C$ = (3.97$\pm$0.07)$\times$10$^{-7}$ emu K/g.  
From the value of $\chi_{\infty}$, we calculate 
the normal-state density of states at the Fermi level 
to be  2.94 $\times$10$^{-3}$/eV atom.  Since the Fermi energy 
$\epsilon_{F}$ = $-$55 meV and the average diameter of the outermost shells is about 
10 nm \cite{Chau}, the Fermi level is crossing the first subband of nearly 
all the metallic shells.  Assuming 80$\%$
metallic chirality shells and using a theoretical result 
\cite{Mintmire}, we find that the density of states $N(\epsilon_{F}) = 
0.8 (2\sqrt{3}/\pi^{2})(1/\gamma_{\circ})(a_{C-C}/\bar{d})$, where 
$\bar{d}$ is a diameter averaged for all the shells, $a_{C-C}$ = 0.142 
nm and $\gamma_{\circ}$ = 2.2 eV.  If we take the average diameter of 
the innermost shells to be 2 nm and the average diameter of the 
outermost shells to be 10 nm (Ref.~\cite{Chau}), then $\bar{d}$ = 6 
nm.  Substituting $\bar{d}$ = 6 nm into the above equation yields 
$N(\epsilon_{F})$ = 2.86 $\times$10$^{-3}$/eV atom, in excellent 
agreement with that deduced from Fig.~\ref{ESR}.  Moreover, the 
carrier density is $-\epsilon_{F}N(\epsilon_{F})$ = 
1.6$\times$10$^{-4}$/atom = 1.6$\times$10$^{19}$/cm$^{3}$, in 
remarkably good agreement with the measured one 
(1.6$\times$10$^{19}$/cm$^{3}$) \cite{Baum}. ~\\
~\\
{\bf 6~~Meissner effect and Remnant Magnetization in MWNTs} 
\vspace{0.5cm} 

It is well known that the diamagetic susceptibility of graphite is 
very small when the magnetic field is along the plane.  It is also 
shown that the diamagetic susceptibility is small when the field is 
along the tube axis direction.  If we take $d$ = 6 nm and 
$\epsilon_{F}$ = $-$55 meV (see above) and use the theoretically 
calculated result of Ref.~\cite{Lu}, we estimate $\chi_{\parallel} 
(0)$ $\simeq$ $-1.7\times 10^{-6}$ emu/g, which is a factor of 7 
smaller than the measured one ($-1.1\times 10^{-5}$ emu/g) 
\cite{Chau}.  This large discrepancy can be naturally resolved if the 
MWNTs are superconductors that contribute to diamagnetism due to the 
Meissner effect.

For tubes of radius 
$r$ with magnetic field parallel to the tube axis direction, the 
zero-temperature diamagnetic susceptibility due to the Meissner 
effect is given by
\begin{equation}\label{dia}
 \chi_{\parallel} (0) = 
-\frac{\bar{r^{2}}}{32\pi\lambda_{\theta}^{2}(0)}.
\end{equation}
Here $r$ is the 
radius of tubes, $\bar{r^{2}}$ is the average value of $r^{2}$, and 
$\lambda_{\theta} (0)$ is the penetration depth when the screening current 
is along the circumferential direction.  The above equation is valid only if 
$\lambda _{\theta} (0)$ is larger than the maximum radius of tubes.  If we 
assume that the radii of tubes are equally distributed from 0 to 
100~\AA, we find $\bar{r^{2}}$= 6000~\AA$^{2}$.  With the weight 
density of 2.17 g/cm$^{3}$ \cite{Qian} and $\chi_{\parallel} (0) = 
-0.93\times 10^{-5}$ emu/g \cite{Chau} (we have subtracted the 
normal-state diamagnetic susceptibility), we calculate $\lambda_{\theta} (0)$ 
$\simeq$ 1724 \AA.  This value of the penetration depth corresponds to 
$n/m_{\theta}^{*}$ = 0.96$\times 10^{21}/$cm$^{3}m_{e}$, where 
$m_{\theta}^{*}$ is the effective mass of 
carriers along the circumferential direction.  With $n$ = 1.6$\times 
10^{19}/$cm$^{3}$ (Ref.~\cite{Baum}), we find that $m_{\theta}^{*}$ 
$\simeq$ 0.017 $m_{e}$, which is larger than the in-plane effective 
mass of 0.012 $m_{e}$ for graphites \cite{Bayot}.  Since the effective mass along 
the tube axis direction is nearly zero \cite{Paul}, the 
value of $m_{\theta}^{*}$ indicates that MWNTs are quasi-1D 
superconductors.
\begin{figure}[htb]
\ForceWidth{7cm}
\leftline{~~~~~~\BoxedEPSF{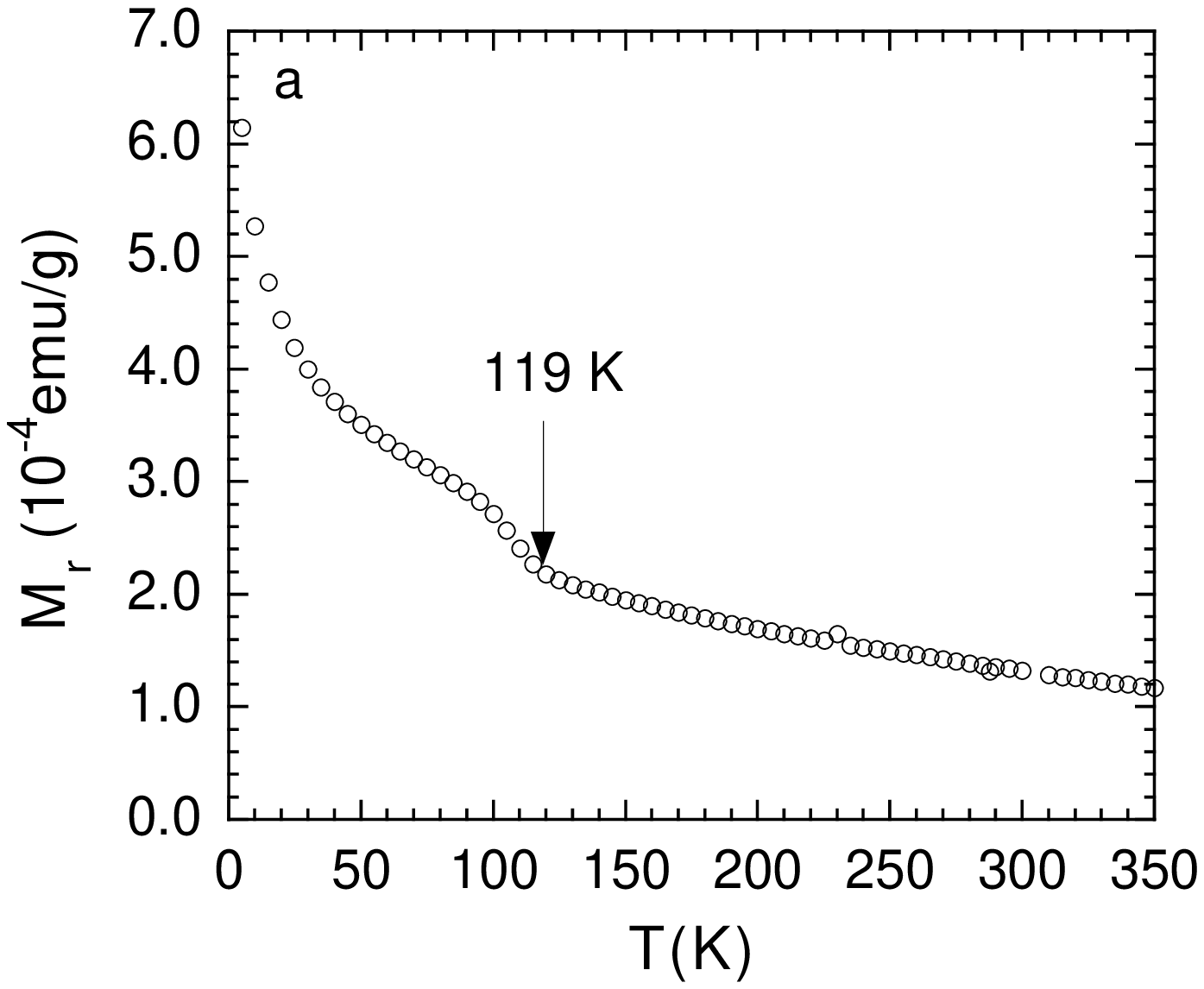}} 
\vspace{-5.7cm}
 \ForceWidth{7cm} 
 \rightline{\BoxedEPSF{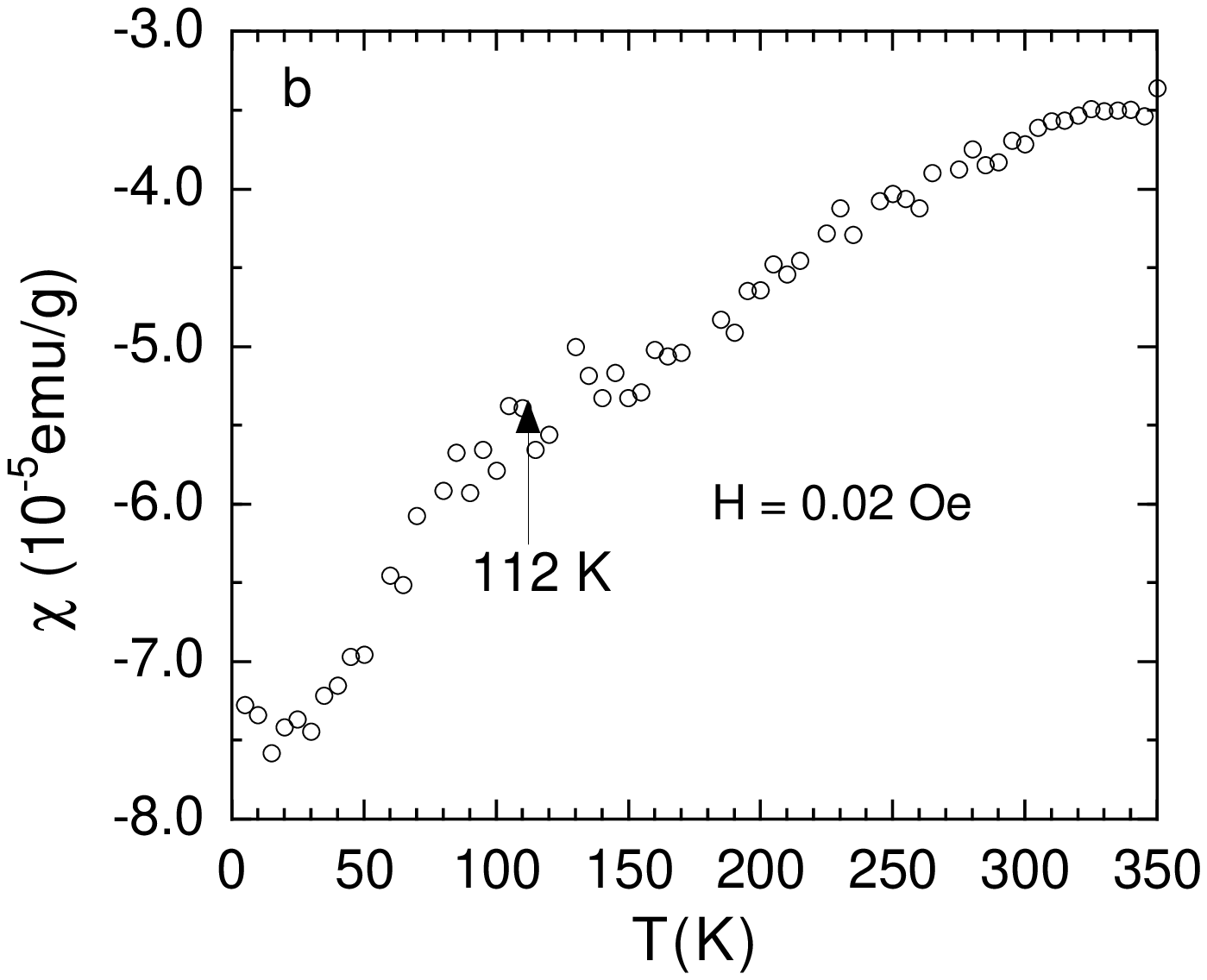}~~~~~~} 
\caption [~]{a) Temperature dependence of the remnant magnetization 
for multi-walled nanotubes.  b) The field-cooled susceptibility as a 
function of temperature in a field of 0.020 Oe.  After 
Ref.~\cite{Zhao}.}
\label{M1}
\end{figure}

From Eq.~\ref{dia}, we can see that $\chi_{\parallel} (0) $ will 
increase linearly with increasing $\bar{r^{2}}$.  For Josephson 
coupled MWNT bundles in unprocessed ropes, the effective $\bar{r^{2}}$ is 
much larger than that for physically separated MWNTs.  This can 
naturally explain why the total low-temperature diamagnetic 
susceptibility for unprocessed MWNTs is larger than that for 
physically separated MWNTs by a factor of 4.3 (see the result of 
Ref.~\cite{Chau}).  Without 
superconductivity in these MWNTs, it is very difficult to account for 
such a large enhancement in the diamagnetic susceptibility for the 
physically coupled MWNTs.

To further show that MWNTs are room-temperature superconductors, 
we show in Fig.~\ref{M1} the temperature dependencies of the remnant 
magnetization $M_{r}$ and the diamagnetic susceptibility for our 
MWNT ropes.  It is apparent that the temperature dependence of $M_{r}$ 
(Fig.~\ref{M1}a) is similar to that of the diamagnetic susceptibility 
(Fig.~\ref{M1}b) except for the opposite signs.  This behavior is 
expected for a superconductor.  Although the $M_{r}$ was also observed by 
Tsebro {\em et al.} up to 300 K \cite{Tsebro}, the 
observation of $M_{r}$ alone does not give unambiguous evidence for 
RT superconductivity since such a $M_{r}$ could be caused by ferromagnetic 
impurities.

We now rule out the existence of ferromagnetic impurities. If there were 
ferromagnetic impurities, the 
total susceptibility would tend to turn up 
below 120 K where the $M_{r}$ increases suddenly. This is because 
paramagnetic susceptibility and $M_{r}$ should increase simultaneously for ferromagnetic impurities.  In 
contrast, the susceptibility suddenly turns down rather than turns up below 120 K 
(Fig.~\ref{M1}b).  This provides strong evidence that the observed $M_{r}$ in 
our MWNTs are not associated with the presence of random ferromagnetic 
impurities but with superconductivity.  Moreover, the large anisotropy 
in the $M_{r}$ of a MWNT rope \cite{Tsebro} suggests that this $M_{r}$ 
is unlikely to arise from random magnetic impurities, because such an 
effect should be isotropic with respect to the field orientation.
~\\
~\\
{\bf 7~~Conclusion} 
\vspace{0.5cm}

Although we may feel that RT superconductivity in carbon nanotubes 
{\em is} too wonderful to believe, we find that the over twenty 
arguments from Refs.  [7-11] and herein are too compelling to be false.  
The mechanism for RT superconductivity may arise from strong 
electron-phonon and electron-plasmon coupling in the quasi-1D 
electronic systems \cite{Lee}.  ~\\
~\\
{\bf Acknowledgment:} I thank Dr.  Pieder Beeli for his valuable comments.  ~\\
 ~\\  
$^{*}$Correspondence should be addressed to gzhao2@calstatela.edu.

\end{document}